\documentclass[12pt]{article}
\usepackage{amssymb,amsmath,amsfonts,graphics,graphicx}
\newcommand{\gsim}{\lower.7ex\hbox{$\;\stackrel{\textstyle>}{\sim}\;$}}
\newcommand{\lsim}{\lower.7ex\hbox{$\;\stackrel{\textstyle<}{\sim}\;$}}

\newcommand{\TeV}{\,\mathrm{TeV}}
\newcommand{\GeV}{\,\mathrm{GeV}}
\newcommand{\MeV}{\,\mathrm{MeV}}

\newcommand{\half}{\frac{1}{2}}

\newcommand{\MM}{\mathcal{M}}
\newcommand{\OO}{\mathcal{O}}

\begin{document}
\begin{titlepage}
\renewcommand{\thefootnote}{\fnsymbol{footnote}}
\setcounter{footnote}{0}
\begin{flushright}
SLAC-PUB-11302\\ 
hep-ph/0506242
\end{flushright}
\vspace{.3in}
\begin{center}
{\Large \bf  Stopping Gluinos}

\vspace{0.5cm}

{\bf 
 A. Arvanitaki$^1$, S. Dimopoulos$^1$\footnote{The work of AA, SD, SR and JW is supported by National Science Foundation grant PHY-9870115 and the Stanford Institute for Theoretical, Physics},\\ A. Pierce$^{2,1}$\footnote{The work of AP is supported by the U.S. Department of Energy under contract 76SF00515.}, S. Rajendran$^1$, and J. Wacker$^1$}

\vspace{.5cm}

{\it $^1$ Physics Department, Stanford University, \\ Stanford,
California 94305, USA}

\vspace{0.2cm}

{\it $^2$ SLAC, Stanford University, \\ Menlo Park, California 94309, USA}

\end{center}
\vspace{.8cm}

\begin{abstract}
Long lived gluinos are the trademark of split supersymmetry.  They 
form R-hadrons that, when charged, efficiently lose energy in matter 
via ionisation.
Independent of R-spectroscopy and initial hadronization, a fraction of 
R-hadrons become charged while traversing a detector. 
This results in a large number of stopped gluinos  
at present and future detectors.  For a 300 GeV gluino, $10^6$ will stop 
each year in LHC detectors, while several hundred stop in detectors during 
Run II at the Tevatron.
The subsequent decays of stopped gluinos produce distinctive depositions  
of energy in calorimeters with no activity in either the  
tracker or the muon chamber.   The gluino lifetime can be determined by looking
for events where both gluinos stop and subsequently decay.
\end{abstract}

\bigskip
\bigskip

\end{titlepage}

\section{Introduction}
\renewcommand{\thefootnote}{\arabic{footnote}}
\setcounter{footnote}{0}

The emergence of the landscape supports the idea 
that the smallness of the observed vacuum energy is explained by a fine-tuning.
The hierarchy problem, historically the primary motivation for new weak-scale physics, might be solved by a similar fine-tuning.  
In this case, new physics is motivated by other considerations, such as gauge coupling unification and the existence of dark matter.  Split superymmetry \cite{NimaSavas,Gian,Wells} is such a concrete implementation of fine-tuned physics beyond the standard model, and will be tested at future colliders.  

Split supersymmetry has a long lived gluino whose lifetime is determined by the 
extent to which the weak scale is fine-tuned.  The gluino is the only 
new TeV-scale colored particle in split supersymmetry (SUSY), and 
therefore is the only particle copiously produced at hadronic accelerators.  
Discovering a long lived gluino is challenging because it typically
leaves detectors without despositing a significant amount of its energy in detectors.
Instead of decaying through a cascade of strong and electroweak transitions, 
the gluino may behave 
much like the lightest supersymmetric particle (LSP), revealing its 
presence through an excess of missing energy events.  
  
If gluinos live longer than tens of nanoseconds, most pass through 
detectors without decaying.  
Gluinos that are stable on detector time scales have motivated several studies of 
split SUSY phenomenology \cite{Kilian,HewettLillieRizzo,Aafke,Cheung:2004ad}.
There are also earlier studies where the gluino is the LSP or NLSP that have
similar phenomenology \cite{Farrar,Gunion,Heister:2003hc,RunII}.
While these searches can find a new, long-lived strongly interacting particle,
they will not determine if this particle is absolutely stable.
In-flight decays of the gluino result in displaced vertices, which indicate a finite 
gluino lifetime, but are only relevant for squark masses between $10^2 \TeV$  
and $10^4 \TeV$ \cite{NimaSavas,RunII,Toharia}. 
There have also been discussions of detecting nearly stable gluinos 
in cosmic ray showers \cite{Cosmics} and if they were discovered
in this fashion, would indicate very long gluino lifetimes. 

In this letter, we point out that an observable fraction of gluinos may 
stop within a few meters of material, and thus within detectors\footnote{
This is analogous to an idea 
studied recently \cite{Feng} for stopping slepton ``next-to lightest supersymmetric particles'' (NLSPs) and examining their late decays. }.
This might allow observation of decays with much longer lifetimes -- up to 
the running time of an experiment. 
For gluino masses greater than about 500 GeV, big bang nucelosynthesis constrains
the gluino lifetime to be less than 100 seconds \cite{Stanford} and motivates searching for late decaying stopped gluinos.
In concert with other ``stable'' gluino signatures, this method 
could play an important role in discovering split SUSY.

The organization of the paper is as follows.  
In Sec. \ref{Sec: Prod} we discuss the production of gluinos at the Tevatron
and the LHC.  These long-lived gluinos hadronize upon production 
into ``R-hadrons''.  In Sec. \ref{Sec: Spec} we discuss the spectroscopy 
and relative production fraction of these particles.  
Even if the majority of R-hadrons are born neutral, we show that a 
population of charged R-hadrons is induced by interactions within 
the detector.   
In Sec. \ref{Sec: Prop} we discuss how this occurs, as we study 
the propagation of the R-hadrons through
matter.  
We then compute the number of stopped R-hadrons as a function of distance 
in iron.  This allows an estimate of the number of stopped gluinos in current 
and future detectors (Sec. \ref{Sec: Stopped}).  When a gluino ultimately 
decays, it will give rise to jets, not originating from the primary 
interaction point but from its stopping point. A detector will observe 
out-of-time energy deposition in the calorimeters which will appear 
as missing energy relative to the interaction region. We discuss these features
in Sec. \ref{Sec: Decay}.   In addition, it is possible to measure the lifetime of
the gluino in events where both produced gluinos stop and subsequently decay 
inside the detector within a relative time of the order of the gluino lifetime.  
When a single produced gluino stop, the second stops about 20\% of the time at
the LHC and 30\% of the time at the Tevatron.  These correlated "double bang" events
are the most promising method 
for determining a long lived gluino lifetime at colliders.

\begin{figure}[t]
\begin{center}
\label{Fig: CrossSection}
\includegraphics[width=4in]{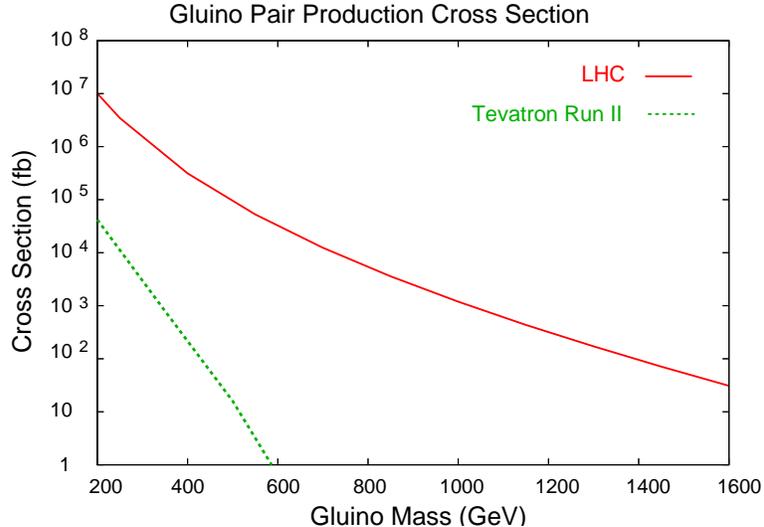}
\caption{The gluino production cross section as a function of mass at the LHC (red solid) and Tevatron Run II (green dashed).}
\end{center}
\end{figure}

\section{Gluino Production}
\label{Sec: Prod}

We calculate the production rate of gluinos at both the LHC
($\sqrt{s} =14$ TeV) and the Tevatron ($\sqrt{s} =1.96 \TeV$) using the
CTEQ4l parton distribution functions (PDFs) \cite{CTEQ}.  To take into 
account the enhancement found at next-to-leading order(NLO), we evaluate 
the leading order expression (see, e.g., \cite{LO}) at $Q^{2}=  (0.2 m_{\tilde{g}})^{2}$, 
where the leading-order and NLO results match \cite{NLO}.  At low masses, the 
gluino production rate at the LHC is extraordinary, 
reaching $\sim$1/sec for $m_{\tilde{g}} \approx 350$ GeV. At low velocities, 
the Sommerfeld resummation of the ``Coulomb ladder'' gives 
an $\pi\alpha_s/v$ enhancement for the production of slow gluinos.  This is particularly relevant 
for the $gg \rightarrow \tilde{g} \tilde{g}$ subprocess where the 
gluons are in an attractive state. We model the Sommerfeld enhancement by 
multiplying the cross section for the $gg \rightarrow \tilde{g} \tilde{g}$ subprocess by
\begin{eqnarray}
E_s =  \frac{C \pi \alpha_s/v }{1 - \exp(- C \pi \alpha_s/v)},
\end{eqnarray}
with $C=1/2$ \cite{Gunion,Brodsky}.
This coefficient comes from a color-averaging of the various initial and 
final states.  Most of the difference between the leading order 
and NLO production
cross sections arises from physics at distance scales much shorter
than those responsible for the Sommerfeld enhancement.   So, it seems
reasonable to treat these two contributions as factorizable. As an
approximation, we take:
\begin{equation}
\sigma = E_{s} \times\sigma_{LO} \Big|_{\mu=0.2 m_{\tilde{g}}} .
\end{equation}
The integrated cross sections for gluino pair production at the
Tevatron and the LHC are shown in in Fig. 1.  We have placed the 
most minimal of cuts,  $|\eta_{\tilde{g}}| < 4$.

\begin{figure}[t]
\begin{center}
\label{Fig: VelocityDist}
\includegraphics[width=3in]{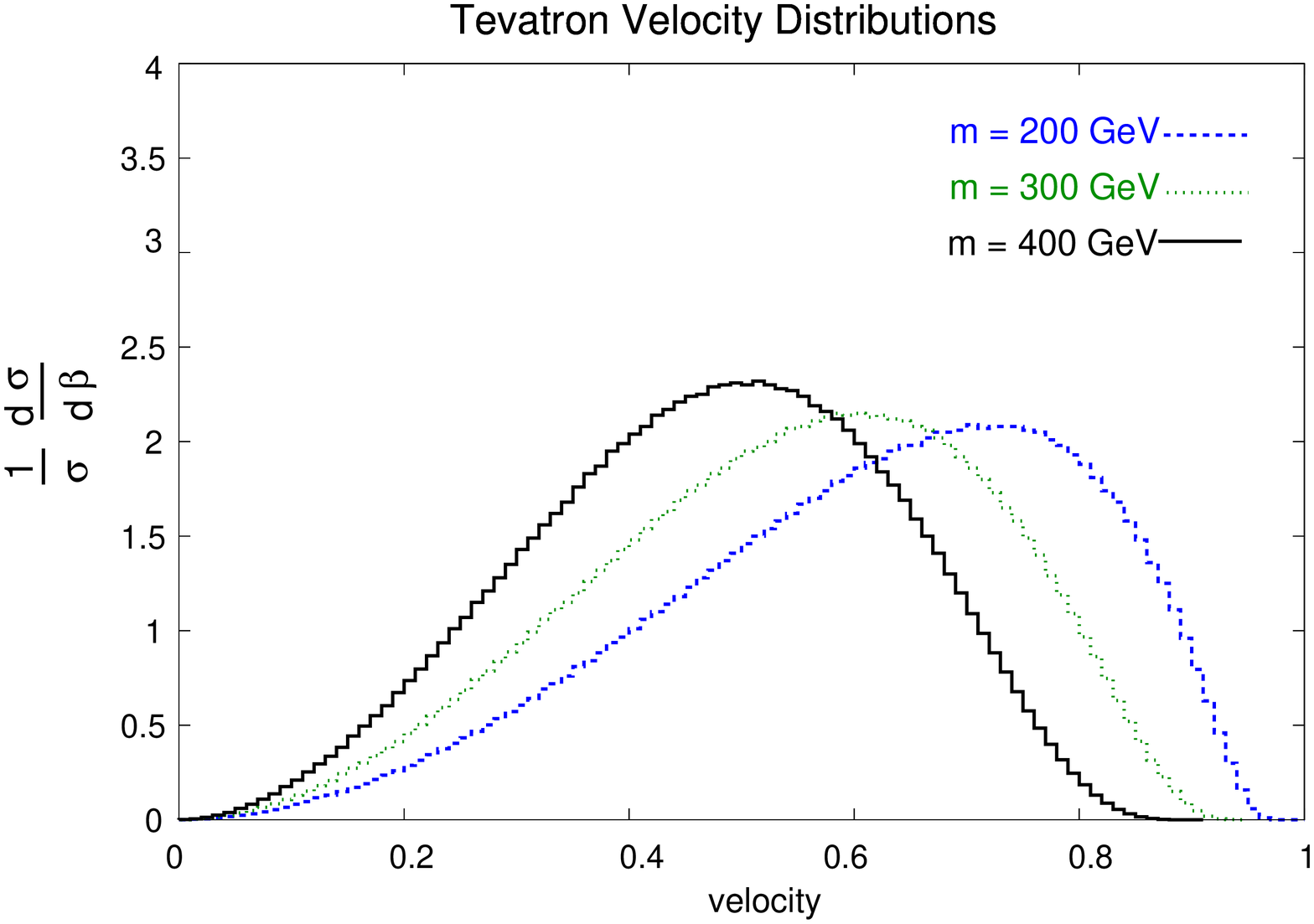}
\includegraphics[width=3in]{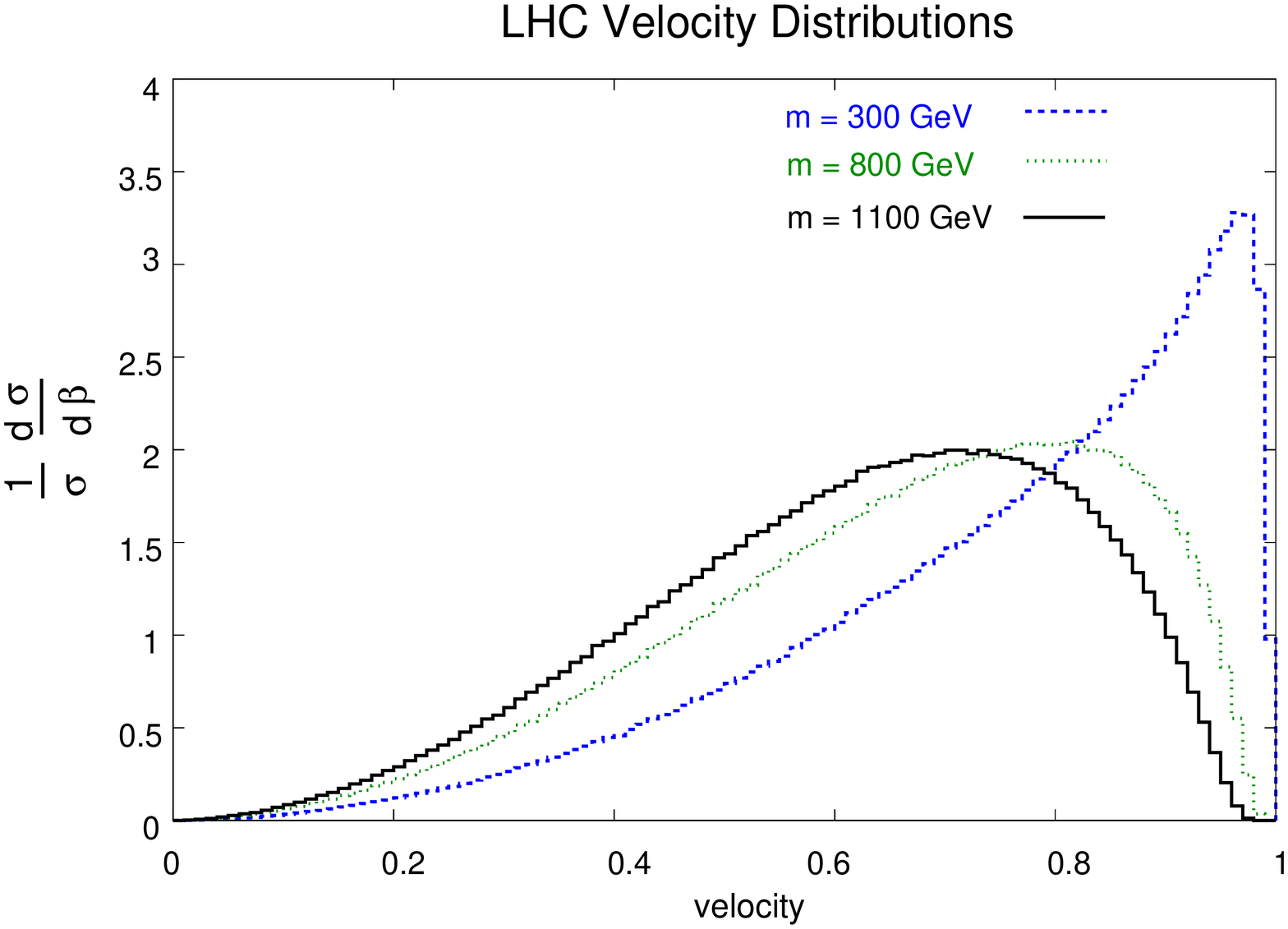}
\caption{The distribution of gluino velocities at the LHC (right) and
Tevatron (left).  In each case we have shown the distribution for multiple
gluino masses.  At the Tevatron, we show $m_{\tilde{g}} = 200, 300,400$ GeV as 
dashed blue, dotted green, and solid black, respectively. At the LHC, we 
show the distribution for $m_{\tilde{g}} =300, 800, 1100 $ GeV as dashed 
blue, dotted green, and solid black, respectively.}
\end{center}
\end{figure}

While the cross section for the gluino production is a steeply
falling function of the gluino mass, the number of slowly moving
gluinos does not fall quite as steeply.  This is because the velocity
distribution skews toward smaller velocities as the mass of the gluino
increases.  We show the normalized velocity distribution in 
Fig. 2.  Even for the lightest masses we consider at the Tevatron, $m_{\tilde{g}}=200 \GeV$, 
the gluino is produced with non-relativistic velocities.    
In contrast, a 300 GeV gluino at the LHC produced
relativistically.  Gluinos at the LHC do not become non-relativistic until masses 
around 1 TeV.

These distributions change as a function of
the pseudo-rapidity, $\eta$.  There are fast gluinos in the forward region due 
to a boost going from the parton center of mass frame to the lab frame.
This trend can be seen in Fig. 3 for a 300 GeV gluino at the LHC.  
While somewhat moderated at higher masses, the trend persists.  
Since only the slowest gluinos will stop, the stopped 
gluinos will preferentially be in the central part of the detector.  
We revisit this point in Sec.~\ref{Sec: Stopped}.

\begin{figure}[h]
\label{Fig: Rapidity}
\begin{center}
\includegraphics[width=3.5in]{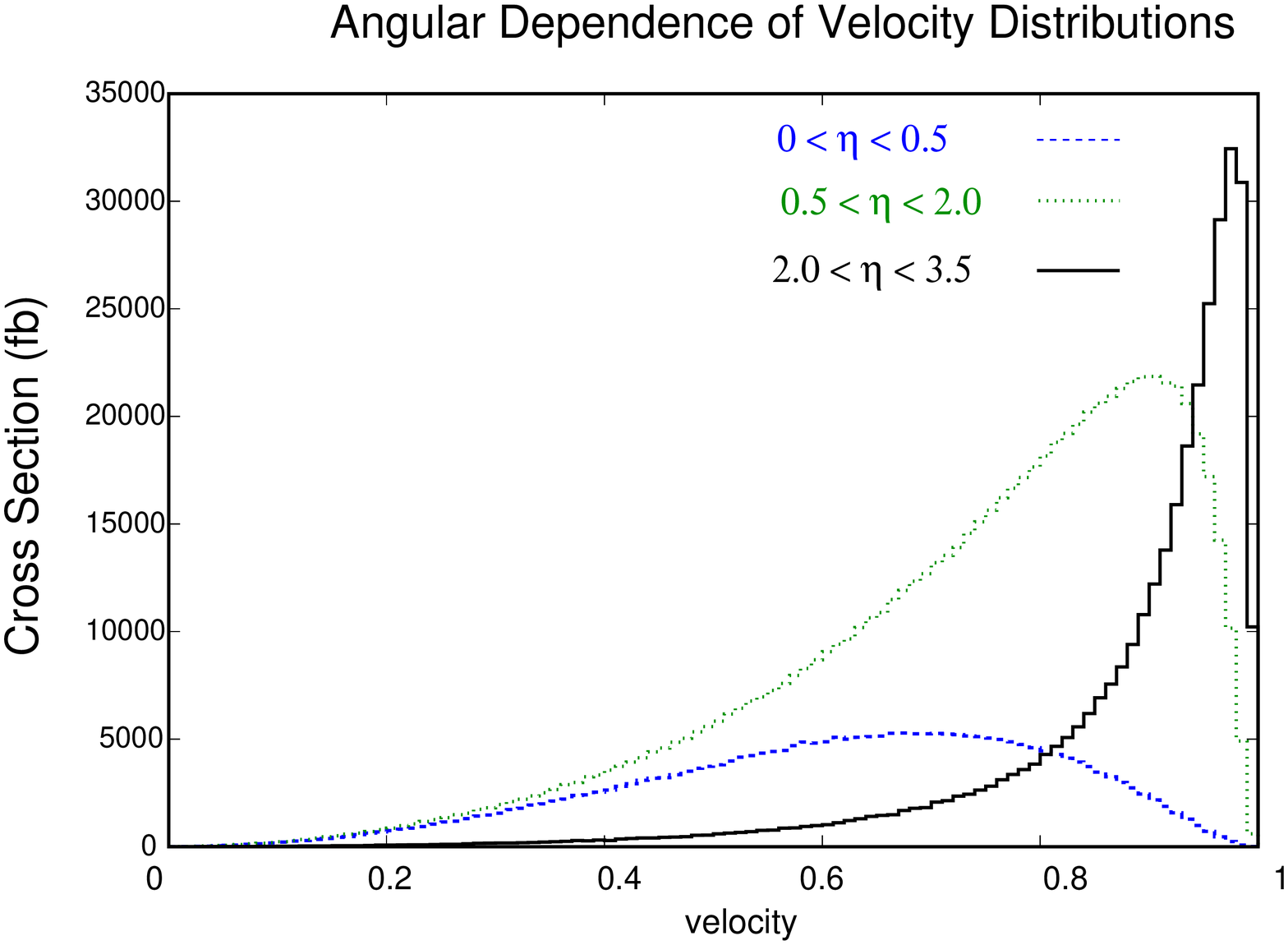}
\caption{The distribution of gluino velocities at the LHC for 
$m_{\tilde{g}}=300$ GeV as a function of rapidity.  Shown are curves for 
$0 < |\eta| < 0.5$ (dashed blue ), $0.5 < |\eta| < 2.0$ (dotted green), 
$2 < |\eta| < 3.5$ (solid black).}
\end{center}
\end{figure}

\section{Spectroscopy and Hadronization}
\label{Sec: Spec}

After production, gluinos combine with light degrees of freedom to 
form colorless hadrons. The mass spectrum of these hadrons will affect the 
propagation of gluinos through the detector.  We do not compute the 
spectrum of the R-hadrons from first principles.
Rather, we enumerate the relevant possibilities, and look at the 
consequences of each.   Electromagnetic interactions will be the dominant
mechanism for R-hadrons to stop.  We therefore pay particular attention to whether there are 
long lived charged R-hadrons.

We will treat isospin as a good symmetry since the dominant
breaking arises through light quark masses and electromagnetic effects.  The 
splittings in the isospin multiplet are a few MeV and for the
energies and processes we are considering, these splittings are sufficiently
small to be irrelevant.  For simplicity, we exclude the possibility of 
strange valence quarks. Including them would not significantly affect 
the results of this paper.

\subsubsection*{Mesons}

R-mesons can be ($\tilde{g} g$) or ($\tilde{g} q \bar{q}$) states.  
The ($\tilde{g} q \bar{q}$) states can be classified by their isospin.  
Considering only $u$ and $d$ quarks, there is an isosinglet and an 
isotriplet.

The R-meson spectroscopy is the most uncertain because we know
very little about the constituent mass of a gluon from QCD.  
The constituent mass of the gluon has been estimated to be 
700 MeV, but there are large uncertainties.  Because it is roughly about
twice the constituent mass of the a light quark, the
mass ordering of a gluino-gluon state ($R_{g}$) and the   
gluino-quark-anti-quark state ($R_{q\bar{q}}$) is unclear.  Moreover, 
the isosinglet $R_{q\bar{q}}$ will mix with $R_g$ and the distinction 
between these two states is artificial.  

The isotriplet possesses charged states which will efficiently lose
energy, so it is important to determine how often the gluino is in the 
isotriplet, and how often it is in an isosinglet state.
Thus, the most important aspect of R-meson spectroscopy is
the mass of the lightest isotriplet state  relative to the lightest
isosinglet state.  We denote this mass difference by 
$\Delta M_{31} \equiv M_{\mathbf{3}} - M_{\mathbf{1}}$.   

There are three distinct possibilities for $\Delta M_{31}$: 
\begin{itemize}
\item Mass Region 1: $\Delta M_{31}>m_\pi$ \hspace{0.60in}  Isosinglet is long lived.
\item Mass Region 2: $|\Delta M_{31}|<m_\pi $ \hspace{0.5in} Both are long lived.
\item Mass Region 3: $\Delta M_{31}< -m_\pi$ \hspace{0.5in} Isotriplet is long lived.
\end{itemize}
In the first and last case, the heavier meson will be unstable to 
strong decays.  Thus, all mesons exiting the interaction region
will either be isosinglets or isotriplets, respectively.   In Mass Region 1, all 
R-mesons exiting the interaction region will be neutral.  In Mass Region 3, 
two thirds will be charged.  These represent 
two extreme cases; the middle region interpolates between the two.  
As we will see, the final results for stopping will often be  similar even for 
Regions 1 and 3, indicating insensitivity to the spectroscopy.

If the mass splitting $|\Delta M_{31}|$ is less 
than $m_\pi$, the only way for the heavier
state to decay is through a weak transition. 
Weak decays that change the spin by one unit have a mean lifetime of
\begin{eqnarray}
\tau_{\Delta j =1} = \tau_n \left(\frac{ \Delta M_{np}}{\Delta M_{31}}\right)^5 \sim 
60 \text{ ns} \left(\frac{130 \text{ MeV} } {\Delta M_{31} }\right)^5,
\end{eqnarray}
where $\tau_n$ is the neutron lifetime and $\Delta M_{np}$ is the 
proton-neutron mass difference.
This lifetime is of the order of the stopping time of the R-hadron and 
therefore is not important qualitatively.  In particular, if the 
available phase space is even somewhat small, then this 
effect can be completely 
neglected.  Decays will occur after a gluino has stopped.  A decay will give 
the gluino a tiny kick, but the velocity will be sufficiently small that the
gluino will rapidly come to rest again.

In Mass Region 2 both the isosinglets and isotriplets live long enough that we  
must consider how both propagate through the detector and make assumptions about their 
initial production fraction.  The produced ratio of isotriplet to isosinglet R-hadrons 
exclusively involving light quarks is easy to estimate:
$N_{R_{\mathbf{3}}}:N_{R_{\mathbf{1}}}= 3:1$.  
The main uncertainty in hadronization is
the number of $R_{g}$ that are made and how light they are relative
to the $R_{q\bar{q}}$  isosinglet.  There is no reason to expect $R_{g}$ production 
to completely dominate the production of the isotriplet state.

\subsubsection*{Baryons}

The next question is whether there are long-lived charged 
R-baryons ($\tilde{g} q q q$).  Since splittings within an 
isomultiplet are small, this question is equivalent to 
whether the lightest baryonic state is an isosinglet or 
a larger isomultiplet.

It seems unlikely that the lightest R-baryon is an 
isosinglet ($\tilde{g} uds$) because it requires the inclusion of a 
strange quark, costing an extra $\sim$150 MeV.  
While this possibility can not be excluded completely, we
temporarily ignore it, and revisit its consequences in
the conclusion. The remaining possibilities are that   
either an isodoublet or isoquartet is the lightest baryonic multiplet.  The 
isoquartet has an intruiging doubly charged state, but for simplicity we 
will assume the isodoublet is the lightest. The conclusions related to 
stopping are insensitive to this assumption.

The most important spectroscopic feature is the difference in mass 
between an R-meson and an R-baryon.  If the 
inequality $M_{R_M} + m_N > M_{R_B} + m_\pi$ is satisfied, then there
are exothermic conversions of R-mesons to R-baryons as the R-hadrons 
propagate through matter (see Sec.~\ref{Sec: Prop}).  This seems 
very likely since the pion is anomalously light due to its 
pseudo-Goldstone nature.  We will assume this inequality throughout 
the paper.  Because there is no strange matter in the detector, matter 
conversion will only produce isodoublet R-baryons and not isosinglets.    

Roughly only $\OO(1\%)$ of the R-hadrons 
produced directly will be R-baryons or R-anti-baryons\cite{Aafke}.  
As we will discuss in the next section, the dominant process for 
producing slow R-baryons is conversion of 
R-mesons in matter, rendering the uncertainty in this
initial hadronization fraction largely irrelevant for determining
the fraction of stopped gluinos.

\section{Propagation Through Matter}
\label{Sec: Prop}

When the R-hadrons are charged, the dominant energy loss is through 
ionization as described by the Bethe-Bloch equation.  Neutral R-hadrons 
will not slow 
appreciably because they have huge momenta and no long range forces.  
We will discuss several interactions important for determining 
the fraction of the time that the R-hadrons are charged.

\subsection{Electromagnetic Energy Loss}

Energy loss via ionization will effectively stop non-relativistic 
charged particles.   As discussed in Sec.~\ref{Sec: Prod}, a 
reasonable fraction of the R-hadrons are slow, particularly in the 
central region.   

The  Bethe-Bloch formula for the rate of energy loss in matter is:
\begin{eqnarray}
\frac{d E}{dx} = - \frac{4\pi  \alpha^2 \rho }{A m_p m_e} \frac{ Zz^2}{v^2} \left(
\half \ln \frac{ 2 m_e v^2  
}{I
(1- v^2)} - v^2 
\right),
\end{eqnarray}
where $E$ is the energy of the incident particle; $A$ and $Z$ are the atomic mass and number of the absorber, respectively;  $\rho$ is the mass density 
of the material; $z$ is the charge of the incident particle;
and $m_e$ and $m_p$ are the masses of the electron and proton respectively.  
In the non-relativistic limit,  $E \simeq m_{\tilde{g}} + \half m_{\tilde{g}} v^2$, and 
the Bethe-Bloch equation can be recast in the form
\begin{eqnarray}
\label{Eqn: BB}
\frac{d v}{dx} = -\frac{1}{ x_0 v^3}\left( 1 + \frac{\log v}{\kappa} + \OO(v^2)\right),
\end{eqnarray}
where $x_0$ and $\kappa$ are material dependent.  
Parametrically these two constants are given by
\begin{eqnarray}
x_0 \sim \frac{1}{\alpha m_e} \frac{m_{\tilde{g}}}{m_e} \frac{1}{4\pi Z \alpha^4\log \alpha^{-1}}
\hspace{0.5in} \kappa \sim \log \alpha^{-1} .
\end{eqnarray}
For iron, setting $m_{\tilde{g}}=500$ GeV, these are $x_0= 526\text{m}$ 
and $\kappa=4.23$ (see Table 1 for other materials).

Insight into the approximate stopping distance can be found 
by dropping the $\log v$ term in the Bethe-Bloch formula. Integration yields:
\begin{eqnarray}
x =  \frac{x_0 v^4}{4} \left(\frac{ m_{\tilde{g}}}{500 \GeV}\right).
\end{eqnarray}
Thus, given a fixed length $x$ of material, all particles with velocities beneath
\begin{eqnarray}
v \lsim \left( \frac{4 x}{x_0}\right)^{\frac{1}{4}} \left( \frac{500\GeV}{m_{\tilde{g}}}\right)^{\frac{1}{4}}
\end{eqnarray}
will stop.  Most detectors have the equivalent of one to two meters of iron in the 
radial direction; therefore, 500 GeV gluinos will stop in detectors if their velocities 
are less than $v \sim 0.30 - 0.35$.

\begin{table}
\begin{center}
\begin{tabular}{|c|c|c|}
\hline
&$x_0$&$\kappa$\\
\hline\hline
Iron&526 m& 4.23\\
Lead& 503 m & 3.60\\
Uranium& 313 m & 3.53\\
Copper& 491 m & 4.10\\
\hline
\end{tabular}
\caption{
\label{Tab: BB}
Coefficients of the Bethe-Bloch equation for common materials in detectors
for $m_{\tilde{g}} = 500 \GeV$ \cite{dedxRefs}.  See Equation \ref{Eqn: BB}.
}
\end{center}
\end{table}

The Bethe-Bloch formula breaks down for velocities beneath the velocity 
where ionization reaches its maximum,  $v_{\text{max ion}}$.  This 
velocity is given by  $v_{\text{max ion}} \simeq \exp(-\kappa+ \frac{1}{3})$.  
Since $\kappa$ is parametrically  
$\OO(\log \alpha^{-1})$, $v_{\text{max ion.}} \sim \OO(\alpha)$.  
Physically, this is when the incident particle becomes 
an adiabatic perturbation on the electrons in an atom.   
For iron, $v_{\text{max ion.}} = 0.015$.  
Stopping below this velocity can be described by the Fermi-Teller theory
\cite{FermiTeller}.  
In Fermi-Teller,  $dv/dx$ is constant for velocities beneath 
$v_{\text{max ion.}}$.  This theory was later extended by
Lindhard and Scharff\cite{LSS}.  We approximate the post-Bethe-Bloch region
beginning at $v_{\text{max ion.}}$. In this approximation the stopping distance 
beneath this velocity is 
\begin{eqnarray}
\Delta x =  3 \kappa\, x_0\; v_{\text{max ion.}}^4  \sim \frac{m_{\tilde{g}}}{\alpha^2 m_e^2} \sim 1 \text{ cm}.
\end{eqnarray}
Extrapolating from the data of muons stopping on copper \cite{Eidelman:2004wy}, 
the stopping distance for a 500 GeV gluino after the break down of the Bethe-Bloch formula is  
$\Delta x = 0.25 \text{cm}$.  This is on par with the above estimate.   

\subsection{Matter Conversion}

The R-hadron may interact with nucleons as it propogates through matter.  
Interactions of the R-hadron
with light degrees of freedom are inefficient at slowing the R-hadron.   
In analogy to a bowling ball moving through a sea of ping-pong
 balls, the kinematics prevent any appreciable momentum loss by the incident 
gluino.  However, these interactions can have an important indirect 
effect--they can cause an uncharged R-hadron to acquire charge.  The 
charged particle can then efficiently lose energy via ionization.  
We will therefore treat these interactions as ``label-changing'' events. 

\subsubsection*{Meson to Baryon Conversion}

We consider a process in which $R$-mesons can interact with baryons in the \
detector to become R-baryons:  $R_{m} + \text{Baryon} \rightarrow R_{b} + \pi$ \cite{Aafke}.
As mentioned in Sec. \ref{Sec: Spec}, this process is likely 
exothermic because the pion is an anomalously light meson 
as it is a pseudo-Goldstone boson.

Exothermic reactions have enhanced cross sections at low velocities. 
This is derivable from Fermi's golden rule:
\begin{eqnarray}
\sigma v_{\text{rel}} \propto  |\MM|^2 \frac{p_f^2}{ v_f} ,
\end{eqnarray}
where $|\MM|^2$ is the matrix element for the transition, $v_{\text{rel}}$ 
is the relative velocity of the incoming state and $p_f^2/v_f$ is the 
result of the final-state density of states.   For the case of a 
light outgoing particle, the density of states simplifies 
to $m_f^2 v_f$.  The cross section then goes as
\begin{eqnarray}
\sigma = \sigma_0 \frac{v_f}{v_{\text{rel}}} .
\end{eqnarray}
For exothermic processes, $v_f$ is fixed; therefore as $v_{\text{rel}}\rightarrow 0$
the cross section increases.  

Thus, at small velocities meson to baryon conversion is enhanced.  After a 
short distance, the R-mesons are depleted and transform into isodoublet R-baryons.  
The exothermic nature of this interaction also ensures
that once R-hadrons are converted to R-baryons, they will not revert to 
R-mesons.  As we will see, the low-velocity enhancement will make 
our results largely insensitive to assumptions about spectroscopy and 
cross sections.

To parameterize our ignorance about the strong interactions, we 
consider three different conversion cross sections, 
$\sigma_{0}=  30\text{ mb}, 3\text{ mb}, 0.3\text{ mb}$.  
In all cases, the cross sections will be enhanced at low velocities by 
the ratio $v_\pi/v_{\text{rel}}$, where $v_\pi$ is the velocity of 
the outgoing pion.  
R-meson to R-baryon conversion should release several hundred MeV
of energy ($Q \approx 400$ MeV), ensuring that the outgoing pions are 
relativistic. We set $v_\pi \simeq 1$ from now on.

The R-meson will scatter off nucleons confined within a nucleus.  These nucleons are not
at rest with respect to the lab frame, and have a Fermi velocity $v_F$.
The average binding energy per nucleon in a nucleus is $\OO(8\MeV)$ 
which means that the Fermi velocity is $v_F\simeq 0.15$.   For scatterings
with an incident velocity less than the Fermi velocity, $v_{\text{rel}} \approx v_F$, while for incident velocities faster than $v_F$ the incident velocity, $v_{\text{rel}} \approx v_{\text{inc}}$.

\begin{table}
\begin{center}
\begin{tabular}{|c|c|}
\hline
&$\ell$\\
\hline\hline
Copper&0.15\text{ m}\\
Iron&0.17\text{ m}\\
Lead&0.17\text{ m}\\
Uranium&0.11\text{ m}\\
\hline
\end{tabular}
\caption{\label{Tab: Nuclear Ints}
Nuclear interaction lengths for common material in calorimeters.
}
\end{center}
\end{table}
The interaction lengths in iron for each of the three benchmark conversion
cross sections are
\begin{eqnarray}
\ell_{\text{Fe}}^{30\text{mb}} = 0.17 \text{ m } v_{\text{rel}}
\hspace{0.3in}
\ell_{\text{Fe}}^{3\text{mb}} = 1.7 \text{ m } v_{\text{rel}}
\hspace{0.3in}
\ell_{\text{Fe}}^{0.3\text{mb}} = 17 \text{ m } v_{\text{rel}}.
\end{eqnarray}
Note, the $30$ mb interaction length corresponds to the interaction length 
of an ordinary pion in material. In Table \ref{Tab: Nuclear Ints} we give the nuclear interaction lengths for other
common materials in detectors.
For the first two cases, the majority of slow R-mesons are
depleted by the end of a few meters, while even for the smallest cross section,
a reasonable fraction of the slowest R-mesons are converted at the end of a typical detector.
Only slow R-hadrons have a chance of stopping in a detector since the stopping distance 
goes as $x_{\text{stop}} \sim x_0 v^4$ with $x_0\sim 500 \text{ m}$.   Therefore, the 
R-mesons that are capable of stopping are converted to R-baryons.    This
insulates us from the details of spectroscopy and even makes the distinction between
a conversion cross section of \mbox{30 mb} and $\mbox{3 mb}$ small.  In fact,
the only situation which will be appreciably different than the others is when the 
isosinglet R-meson is significantly lighter than the 
isotriplet \begin{it}and\end{it} the conversion cross
section is anomalously small.

R-meson to R-baryon conversion is the dominant process for slow R-baryon production.  
Only isodoublet baryons are produced in this process.
We assume that the isosinglet baryon is not several hundred MeV lighter than the 
isodoublet baryons, 
so that the isodoublet baryons are long-lived.
Since the isodoublet has a charged state, these stop efficiently. 
A fraction of the R-mesons will thus be
converted early to R-baryons and then have sufficient time to stop.

\subsubsection*{Charge Oscillation}

Pion exchange will switch the R-hadrons between the states of 
an isomultiplet.  The cross section for this process should be 
comparable to those of strong interactions (30 mb) but could be somewhat 
smaller due to the octet nature of the light quark cloud around the gluino.  
This uncertainty does not significantly affect our results, so we leave
the cross section at 30 mb throughout for simplicity.
With a 30 mb cross section, the interaction length is much shorter than 
the detector. Charge oscillation can therefore be approximated by a
R-hadron that spends a fraction of its life charged.  A doublet 
R-baryon will spend half its time charged, while the isotriplet R-meson will
spend two-thirds of its transit time charged.  During the time they are 
charged, the R-hadrons undergo Bethe-Bloch deceleration, and while 
they are neutral, they propagate freely until their next charge 
exchange process.

In Mass Region 2 ( $|\Delta M_{31}|< m_\pi$) both the isotriplet and 
isosinglet are long lived.  The isosinglet can become an isotriplet 
and vice versa by emitting a pion. We again estimate that this 
cross section is 30 mb.  One of these transitions is exothermic and 
becomes important at low velocities, while the reverse process will be 
endothermic and will turn off at low velocities.     
When the isosinglets are heavier than the isotriplet they are
depleted and the remaining isotriplets will spend $\frac{2}{3}$ 
of the their time charged.    This means that they will stop more quickly
than R-baryons, which only spend one half their time charged. 
So, in the case where the isotriplet is light, a greater 
number of R-hadrons can stop in a detector if the R-meson to R-baryon 
cross section is small.

\subsection{Nuclear Capture}

After R-hadrons have slowed significantly, they may be 
captured by a heavy nucleus.  We view this process as the likely 
final state for a stopped gluino.  Thus even if a stopped 
charged R-hadron transitions to a neutral state, it does 
not ``wander off'' and is really trapped.  

Nuclei capture all hadronic particles provided
they satisfy the following criteria for absorption.  First, the
incident particle should not transfer too much momentum in a single
collision to kick out a nucleon from the absorber.  Second,
the momentum transfer should be small enough so that the incident
particle  couples coherently to an entire nucleon rather than individual
quarks, $\delta q \lsim \Lambda_{\text{QCD}}$.  
Finally, in the center of momentum frame, both particles must come to rest.
In the case we consider here, the kinematics are different than 
typical nuclear physics processes where a heavy nucleus absorbs a 
light incident particle.  Nuclear capture of R-hadrons is near the opposite limit:  
the absorbing nucleus is the light object and in the center of momentum frame the R-hadron 
is nearly stationary.  The condition for capture is that the nucleus must come to rest. 

Particles that have slowed down sufficiently to be captured 
are charged and therefore have isospin.  Thus the stopping R-hadron likely
couples to pions, and we estimate $\sigma \sim 1/ \Lambda_{\text{QCD}}^2$.  
We now consider whether the above criteria are satisfied.  The number of 
interactions in a nucleus scales as $N_{\text{int}} \sim A^{\frac{1}{3}}$.
The condition for absorption becomes
\begin{eqnarray}
\Delta q \sim N_{\text{int}}\delta q = A^{\frac{1}{3}} m_n v_F \gsim A m_n v,
\end{eqnarray}
where $v_F$ is the average Fermi velocity of nucleons within the nucleus.
Now the condition for binding becomes
$v \lsim A^{-\frac{2}{3}} v_F$.   This shows that at slow enough velocities R-hadrons are captured.
The total momentum transfer necessary for the R-hadron to be absorbed is
smaller in lighter nuclei which shows that they are better absorbers of R-hadrons.

\section{Stopped Gluinos}
\label{Sec: Stopped}

To calculate the number of R-hadrons stopped at CDF, D0, ATLAS, and CMS, we 
utilize the initial velocity distributions
from Sec. \ref{Sec: Prod} and incorporate the interactions from 
Sec. \ref{Sec: Prop}. We also characterize how the stopped gluinos 
are distributed
throughout the detectors.  
All the presented results will be for Mass Region 1, where the  
lightest R-meson is an isosinglet.  This represents 
a ``worst-case scenario'', where stopping is due solely to 
conversion to R-baryons.  If there 
are stable charged mesons, as in Mass Region 3, these too would slow down via 
ionization losses, leading to more stopped R-hadrons.  

We consider three cases in this section corresponding to 
R-meson to R-baryon conversion cross sections of  \mbox{$\sigma_{0}= 30, 3, 0.3$ mb}.
There is little difference between the two higher cross sections.  
For either of these two cross sections, R-mesons rapidly convert to R-baryons at the low 
velocities that are important for stopping.

\begin{figure}[h]
\begin{center}
\includegraphics[height=3.5in]{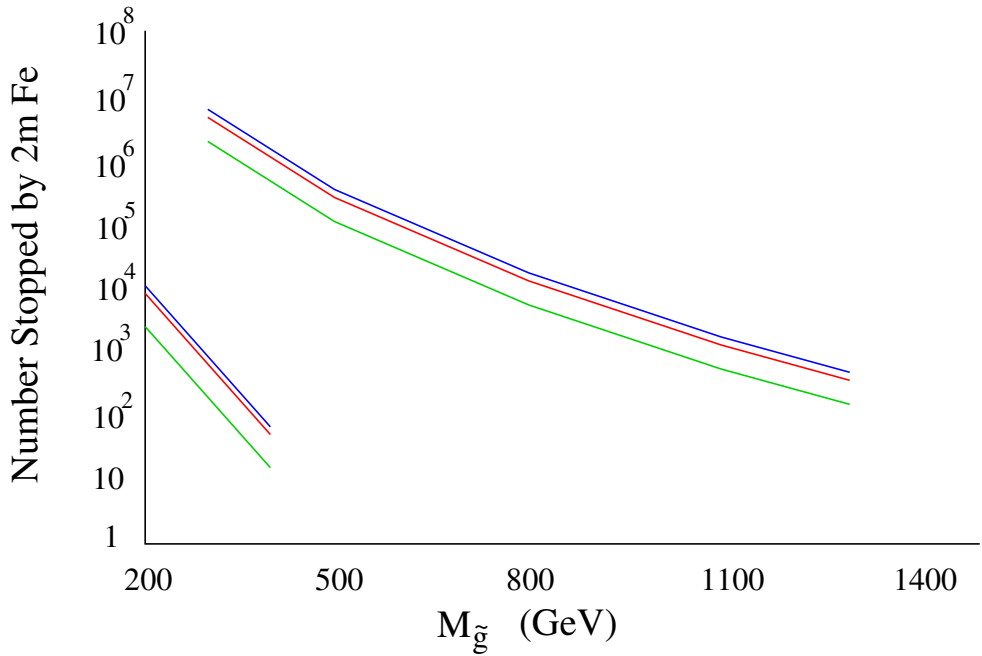}
\caption{
\label{Fig: NStopped}
The number of  R-hadrons stopped after two meters of iron in Mass 
Region 1.  This plot convolutes the velocity distribution at production 
with conversion processes and matter and ionization losses.  The upper set of 
curves is for the LHC for a total accumulated luminosity of 100 fb$^{-1}$, 
equivalent to a year of running at high luminosity. The lower set is for the 
Tevatron Run II, assuming a total of 2 fb$^{-1}$.  In each set the curves correspond to a meson to baryon 
conversion cross section, $\sigma_{0}=$ 30 mb, 3 mb, and 0.3 mb from top to bottom.}
\end{center}
\end{figure}

To estimate the total number of particles stopped, we must estimate 
the total amount of material in each detector.  As a first approximation 
we present the number stopped in 2 meters of iron in Fig. \ref{Fig: NStopped}. 
The difference between the $30 \text{ mb}$ 
and $3 \text{ mb}$
conversion cross sections are relatively small.  
Decreasing the cross section further does make a difference, 
however.  The lowest cross section, $\sigma_{0} = 0.3$ mb, results in an
order of magnitude fewer stopped R-hadrons after 2 meters.

Making a more sophisticated estimate of the number stopped in 
various detectors requires taking into account the detector 
geometries and their different compositions. 
The four main detectors have roughly the following coverage and material 
depths:
\begin{itemize}
\item CDF's EM calorimeter contains 16 cm of Pb in the radial direction.  
The hadronic calorimeter contains 77cm of Fe.  Each calorimeter 
covers  $|\eta|<3.6$.
\item D0's EM calorimeter contains about 7 cm of U; the fine hadronic 
calorimeter has 35 cm of U, while the coarse hadronic calorimeter has 
48 cm of Cu, all 
with coverage out to $|\eta| \lsim 4$.
\item ATLAS's EM calorimeter has 66 cm of Pb, while its hadronic calorimeter contains 156cm of Fe.  Both calorimeters cover up to $|\eta|<3.2$.
\item CMS's EM calorimeter has 46cm of Pb, while its
hadronic calorimeter contains 98 cm of Cu, both cover up to $|\eta|<3$.
\end{itemize}

Taking into account the amount of absorber in each detector, we 
estimate the number of gluinos stopped in Table \ref{Tab: NStopped}.  
We  take the meson to baryon cross conversion cross section to 
be $\sigma_{0} =3$ mb.  The number stopped can be substantial; for instance a 
300 GeV gluino, $\sim 10^{6}$ should stop in each LHC detector in a year of 
high luminosity (100 fb$^{-1}$) running.  At the Tevatron, hundreds of 
300 GeV gluinos stop in each detector after 2 fb$^{-1}$ of running.

\begin{table}
\begin{center}
\begin{tabular}{|c||c|c|c|}
\hline
2 fb$^{-1}$& 200 GeV& 300 GeV & 400 GeV\\
\hline
CDF&$4.1\times 10^3$&$3.1\times 10^2$&$3.3\times 10^1$\\
D0&$4.5\times 10^3$&$3.3\times 10^2$&$3.4\times 10^1$\\
\hline
100 fb$^{-1}$&300 GeV& 800 GeV & 1300 GeV\\
\hline
ATLAS&$5.8\times 10^6$ &$1.8\times 10^4$&$6.2\times 10^2$\\
CMS&$3.7\times 10^6$&$1.2\times 10^4$&$3.9\times 10^2$\\
\hline
\end{tabular}
\caption{\label{Tab: NStopped}
The estimated total number of gluinos stopped in each detector for
a 3 mb R-meson to R-baryon conversion cross section.
}
\end{center}
\end{table}

We now consider the distribution of stopped gluinos within the 
detector.  In Fig.~5, we plot the stopping profiles at both the Tevatron 
and the LHC for propagation through a hypothetical iron detector.  
Curves for different meson to baryon conversion cross
sections are plotted in this figure.  Larger conversion cross sections allow for 
more rapid stopping. 
As can be seen in the figure, for the two larger cross-sections, 
essentially all slow mesons have a chance to convert to baryons.  Therefore,  
a similar fraction stop by 1 meter.  For the smallest conversion 
cross section, not all mesons have converted, and the total number stopped
is substantially fewer.

\begin{figure}[h]
\begin{center}
\includegraphics[width=3.0in]{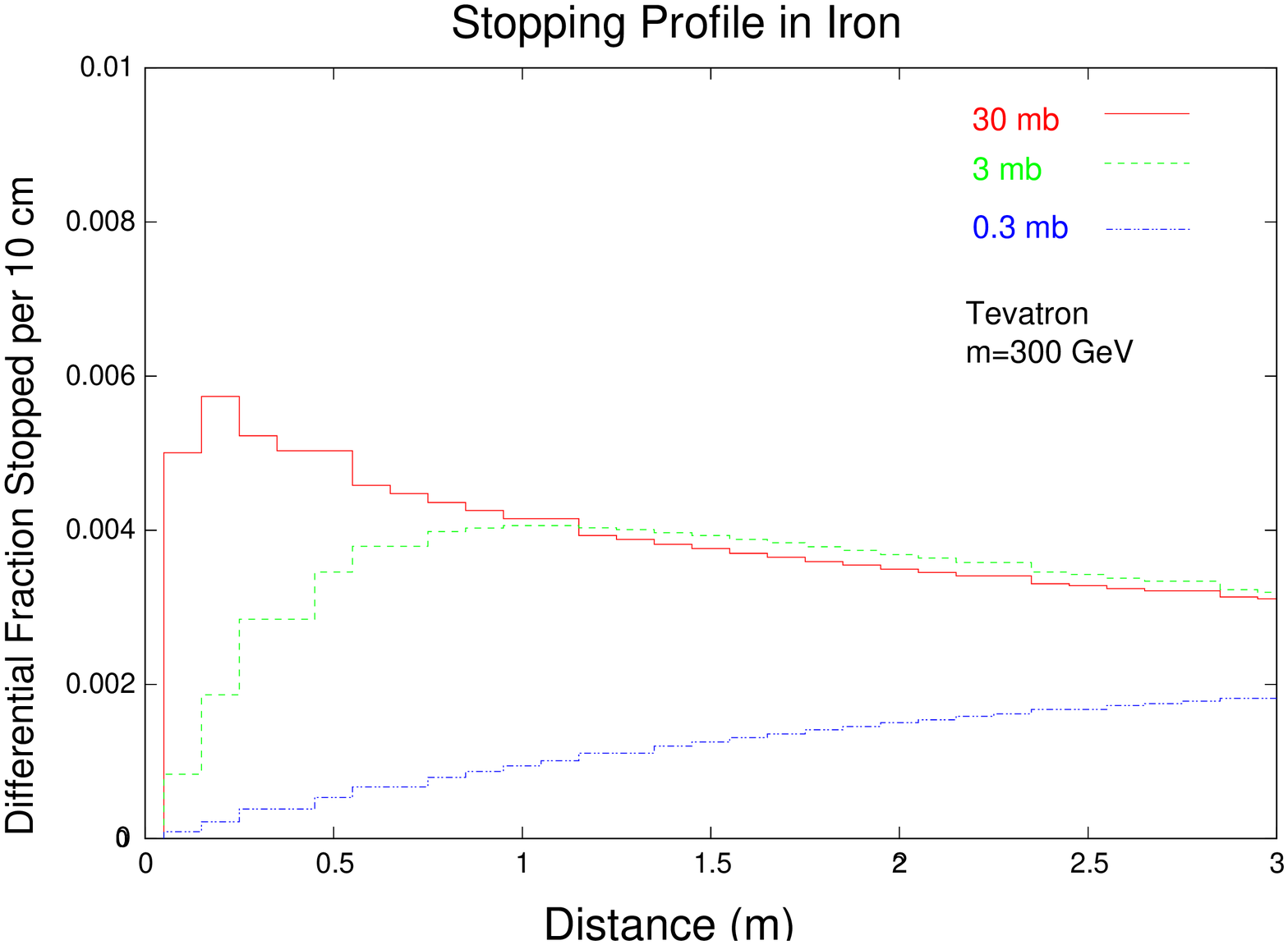}
\includegraphics[width=3.0in]{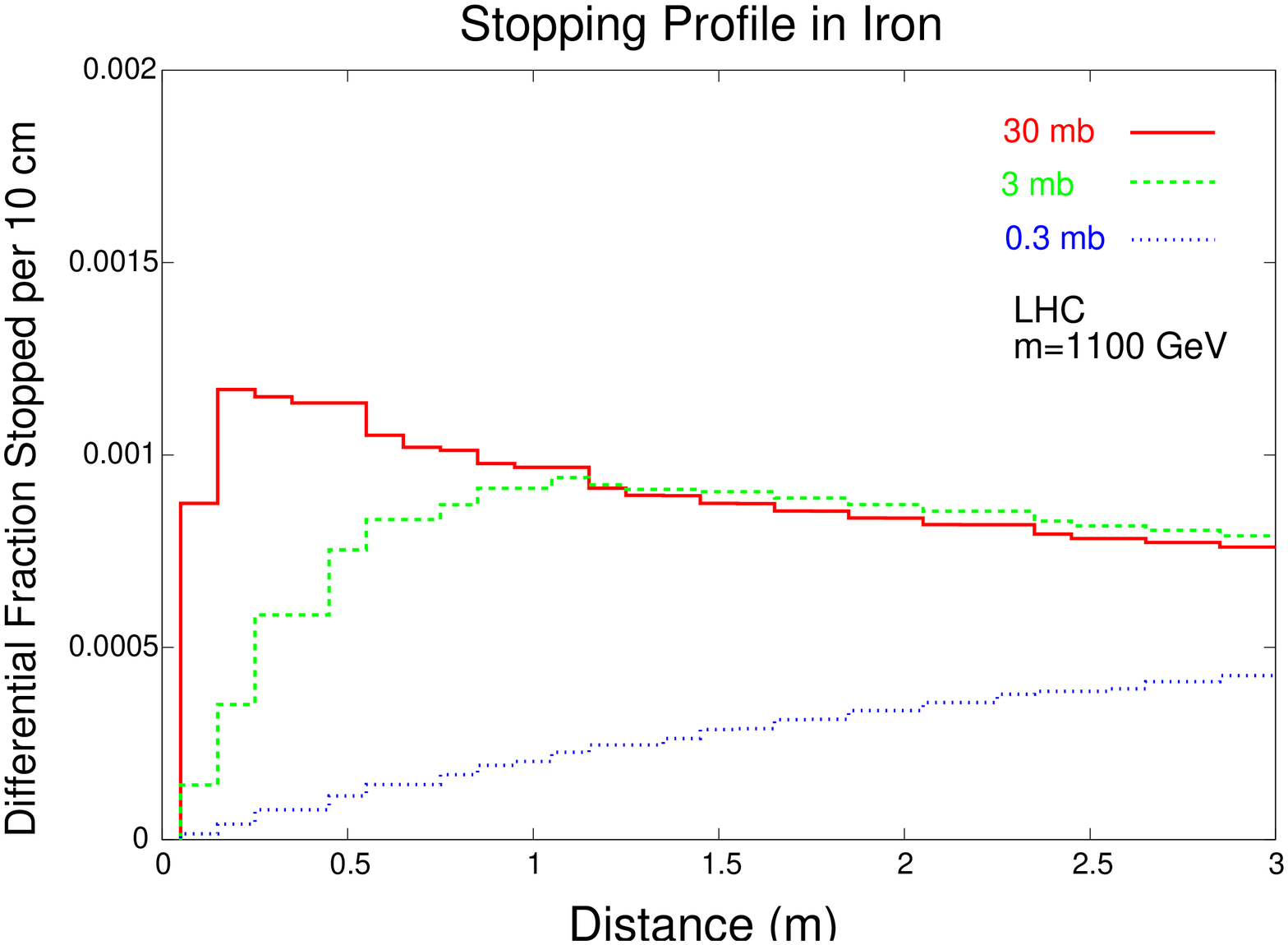}
\caption{
\label{Fig: Profile}
The fraction of gluinos stopped per 10 cm as a 
function of distance in the case where the isosinglet hadron is the 
lightest (Mass Region 1).   There 
are three curves in each plot representing different 
cross sections for R-meson to R-baryon conversion.   The different 
colors represent different cross-sections for meson to baryon 
conversion ($\sigma v_{rel}= 30$ mb (solid blue),
$\sigma v_{rel}= 3$ mb (dashed red), $\sigma v_{rel}= 0.3$ mb(dotted green)).  
The left plot is for the Tevatron and gluino mass of 300 GeV while
the right plot is for the LHC and a mass of 1100 GeV.
} 
\end{center}
\end{figure}

\subsection{Late Decays in Detectors}
\label{Sec: Decay}

A stopped gluino will decay into either a pair of jets and electroweak-ino 
or a single jet and an electroweak-ino.
These jets originate from the point where the gluino stopped and 
will mostly be in the densest
regions of the detector, the electromagnetic and hadronic calorimeters.   The
jets will be pointed in any direction.  
When the decay of the gluino is reconstructed under the assumption that the event came from 
the beam interaction point (as it would be initially), it would appear 
grossly unbalanced.  There would only be activity in the side of the detector where 
the gluino decays.  Therefore, we expect the decays 
to pass the level-one trigger for missing transverse energy.   

The response of the calorimeters to 
jets that are not pointed back to the interaction region is complicated and 
detector dependent.    If the jets are completely contained in a single cell or tower,
this may look like a ``hot cell'' or ``spike.''   A typical 100 GeV jet contains dozens of particles, 
so there will be some leakage into other towers or cells.  
It is beyond the scope of this paper to attempt to calculate how
much leakage there is in each detector, but clearly having several adjacent
cells or towers with energy deposition will help eliminate hardware related 
backgrounds. 

The decay of the gluino happens long after the beam crossing that produced it and  the  
decay will be uncorrelated with any beam crossing. 
At Run II at the Tevatron, the beam crossings are every 396 ns.  
Due  to the relatively long spacing between bunch crossings at
Run II, at CDF 132 ns is recorded around any given bunch crossing.
This means that one third of the decaying gluinos have a chance
to be recorded at CDF. D0 records for the full 396 ns and does not suffer
this efficiency loss. For the ones that are recorded, 
most of the decays will be significantly out-of-time, appearing before or after 
a beam crossing.   This may help reduce QCD backgrounds which appear within 
10 ns of the beam crossing. 
At the LHC, the bunch crossings occur every 25 ns and the calorimeters are
always recording.  This means that any gluino decay can be recorded.
However, the closeness of the bunch spacings means that gluino decays may not
be identified as out-of-time and can not be differentiated from QCD backgrounds
by timing information.

Many gluinos are stopped deep in the hadronic calorimeter and will not 
deposit any energy in electromagnetic calorimeter or in the tracking chamber.    
In the case where the jet is completely contained, the signal is particularly clean.  
The hadronic calorimeter ``lights-up'' with hundreds of GeV of energy, 
but without any activity in any other portion of the detector.  High energy QCD jets
typically have  dozens of charged particles and also deposit energy into
the EM calorimeter and leave tracks.   This seems to indicate that the background
from SM physics is controllable, and the backgrounds seem to be dominated 
by detector backgrounds. 

Backgrounds from cosmic rays are a serious 
consideration since they occur out-of-time 
like a late-decaying gluino.  Most high energy cosmic rays are 
muons because they can effectively penetrate the atmosphere and shielding.  While 
a gluino decays into jets that deposit their energy inside the calorimeter, 
a cosmic ray muon typically transverses the detector without depositing much of its 
energy.  Most cosmic rays will also appear in the muon chamber, so  vetoing muon tracks 
could be a useful discriminant.  Neutral hadronic cosmic rays can not be
vetoed in this way and are a potential background.
However, they do not penetrate matter effectively and when they do reach the
detector, they interact with the outermost layers of the hadronic calorimeter.

\begin{figure}[h]
\begin{center}
\includegraphics[width=\columnwidth]{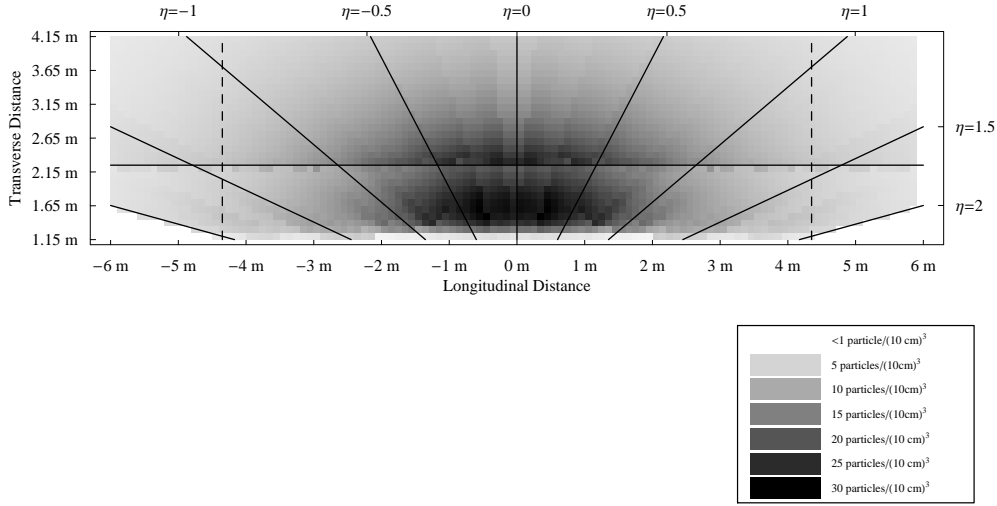}
\caption{
\label{Fig: Atlas}
The density of stopped gluinos in the ATLAS detector 
for $m_{\tilde{g}}= 300$ GeV at the LHC, assuming a 
100 fb$^{-1}$ of data.  Note the increased density of stopped gluinos at low 
pseudo-rapidity.  The vertical lines represent the start of the end cap 
calorimeter.}
\end{center}
\end{figure}
Another unique feature of the signal is that the gluinos that stop are predominantly central
and axially symmetric.  In  Fig. \ref{Fig: Atlas}, we illustrate 
the density of stopped gluinos in the ATLAS calorimeters.  
The density of stopped gluinos can be up to  $30/(10 \text{ cm})^3$ in the central region.  
This distribution distinguishes late decaying gluinos from both 
jet backgrounds (uniformly distributed in rapidity) and cosmic rays (uniform 
throughout the detector and not axially symmetric).  
Depending on the total number of gluinos stopped, 
it might be possible to do sideband subtraction to eliminate QCD and cosmic 
ray backgrounds.

The lifetime of the gluino is set by the supersymmetry 
breaking scale and so is an important quantity to measure.  
It is difficult to determine the lifetime of the gluino directly because it is impossible to
associate the late decay of a gluino with the time of its production.  
Gluinos are pair produced and if both stop, the average  time between the two decays
measures the lifetime of the gluino\footnote{
We thank K. Rajagopal for bringing this point to our attention}. 
For example, consider a 300 GeV gluino with a lifetime of a millisecond.  At 
the LHC, this will result in the production of a pair of gluinos every second, 
and if both stop, there will be two ``bangs'' in the detector 
separated by order a millisecond.  It is not guaranteed that both gluinos will 
stop in any given event.  So long as the R-meson to R-baryon conversion 
cross section is not small, $\sigma_{0} \gsim 1$ mb, we calculate that 
in events where one gluino stops, the second will stop 
20 -- 30\% of the time and therefore a significant number of 
``double bangs'' occur.  If the R-meson to R-baryon conversion cross section 
is smaller, the probability for a double bang depends on the R-spectroscopy and the
initial hadronization of the pair of gluinos.  
This method should be useful in determining the lifetime in all
cases where the decay width is greater than the production rate of stopped gluinos.

A stopped gluino can decay into two jets plus 
a neutralino ($\tilde{g} \rightarrow q \bar{q} \chi^{0}$), or a single jet 
plus a neutralino ($\tilde{g} \rightarrow g \chi^{0}$).  The relative 
branching fraction is sensitive to the gluino mass and the scale of 
supersymmetry breaking\cite{Stanford,Toharia,Gambino}.  The two-body branching 
fraction increases as the gluino mass decreases, or as the scale of SUSY 
breaking increases. It would be of interest to study the extent to which the 
two jets originating within the calorimeter might be disentangled from one
another.  This requires a detailed understanding of the response of the 
detector to jets propagating from within the calorimeter and is beyond 
the scope of this work.  If the two jets can be distinguished, this 
could be an important handle on distinguishing these events from background. 
Even more optimistically, one might get some rough handle on the branching 
ratio of the two-body versus three body decays, thereby indirectly determining
the SUSY breaking scale.

Another possibility is to use the decays of the gluinos into the non-LSP 
electroweak-inos to search for their discovery\footnote{We thank Lian-Tao Wang
for bringing this to our attention.}.  For instance, the 
decay $\tilde{g}\rightarrow \chi_2^0 j\rightarrow \chi_1^0 \mu^+\mu^- j$ will have two energetic 
muons leaving the hadronic calorimeter from the point where the jet deposited its energy.  
This may be a potential discovery channel for electroweak-inos at the LHC.  

\section{Discussion}
\label{Sec: Disc}

Split supersymmetry presents an experimental challenge.  
As in the supersymmetric standard model (SSM), only colored particles are abundantly 
produced at hadronic colliders.  However, unlike the SSM, 
split SUSY does not allow the study of new electroweak particles 
through cascade decays of colored particles -- only the gluino is colored, 
and it decays outside of the inner detector.  
Since electroweak particles are difficult to produce directly, discovering
split SUSY hinges on the identification of the gluino.  
There are now four distinct ways to discover the long-lived
gluino at the Tevatron and the LHC.  
\begin{itemize}
\item 
One possibility is by measuring an excess of monojets
through a $\tilde{g}\tilde{g}j$ event\cite{HewettLillieRizzo}.   
Gluinos will leave very little 
energy in the detector, so the jet will look unbalanced.  However, 
a variety of new physics can lead to a monojet signature.
To claim discovery of split SUSY at the Tevatron or LHC will 
require additional signals beyond monojets.    
It might be possible to distinguish the split supersymmetry monojet 
signature from other monojet events \cite{HewettLillieRizzo}.
It is possible that the gluino might deposit small 
puffs \mbox{($\sim$ 1 GeV)} of 
additional energy as it traverses the calorimeter through hadronic 
interactions.  This would earmark the new physics as strongly interacting.  
\item The second approach is to search for anomalously slow particles in 
the tracking chambers.  
Outside the calorimeters most of the R-hadrons will be R-baryons and therefore
half will be charged.   In \cite{Aafke}, there is a proposal for using the muon chamber 
to search for charged R-hadrons and found a discovery potential for ATLAS up to
1700 GeV.
 $dE/dx$ measurements can also be used in the inner tracking
chamber and look for anomalously heavy charged 
particles \cite{StableCharged}, Reference \cite{HewettLillieRizzo} estimated 
that the Tevatron has a reach of 430 GeV for 
\mbox{2 fb$^{-1}$} if the all R-hadrons are charged and  
a reach of 2.4 TeV at the LHC for \mbox{100 fb$^{-1}$.} 
Unfortunately, in Mass Region 1, the reach of this strategy is reduced 
since almost all of the R-hadrons will be neutral in the tracking chamber. In this case,
only a fraction of the R-baryons will likely be charged, and the reach will 
probably be much closer to 200 GeV for the Tevatron and 1.2 TeV for the LHC.     
\item Another possibility for discovering the gluino is a search for charge
oscillating events, known as ``flippers.'' (see, e.g., \cite{RunII})   
Flippers appear experimentally difficult.  In tracking chambers, where 
charge oscillation can best be measured, there is not much material to 
stimulate the hadronic interactions that lead to the effect.  
\item Seeing a stopped gluino decay is the last known discovery channel
for the gluino.  Only a fraction of the gluinos actually stop, so 
this hurts the reach of this method.  It may take a very careful 
observation of these events to be sure that these are not a fluctuation of a 
QCD event or a cosmic ray.  However, this approach would allow us to
infer the existence of a particle with a finite lifetime.
\end{itemize}
Ideally, some combination of these measurements might lead to a convincing 
discovery of split supersymmetry. 

Since we made several assumptions about strong cross-sections and 
spectroscopy, we would like to comment on the robustness of the 
conclusions derived.  The most robust method of stopping was a two step 
process: R-mesons converted to R-baryons which then stopped via ionization 
losses.  The second part of this process is extremely robust, and relies only
on the well-established physics of the Bethe-Bloch equation.  The first part 
is only slightly more uncertain--it hinges on a reasonable
sized ($\gsim  1/v_{\text{rel}}$ mb) cross section for the meson to baryon conversion.  
If this process were not exothermic, these 
conclusions would not hold.  This seems unlikely given the 
lightness of the pion.  Then, gluinos would only stop if the 
isotriplet mesons are produced, either because they are the lightest 
(Mass Region 3), or by charge exchange (Mass Region 2).  The other 
possible loophole is if the $(\tilde{g} uds)$ state is 
so light that other baryons {\it weakly} decay to it on time scales
faster than the stopping time.  Then meson to baryon conversion only succeeds
in creating   $(\tilde{g} uds)$ hadrons, which do not lose their energy efficiently.  
In this case, stopping gluinos would again rely on having a light 
charged meson.  However, given the mass cost additional strange quark, we view
such a spectroscopy as  unlikely, and consider the conclusions here to  be robust.

We close with a few additional very speculative comments on determining 
the gluino lifetime.  If the gluino is lighter  than 500 GeV, then it is possible for the gluino to
live up to $10^5$ years.  These may never decay inside the detector while
it is running.  Still the R-hadrons will be caught inside the detector.  A search
looking for exotic heavy nuclei could be done in principle, using the 
material of the detector.  In principle, these exotic gluino-containing nuclei 
might even decay at a later time, allowing one to probe lifetimes exceeding 
the duration of the LHC running. 

\vspace{0.5in}
\centerline{\bf{Acknowledgments}}
\vspace{0.2in}

We thank N. Arkani-Hamed for collaboration at an early stage of this work.
We also thank K. Rajagopal for pointing out the ``double bang'' events as
the best way to measure the lifetime of the gluino.
We acknowledge H. Frisch and D. Toback for discussions of calorimeter 
response and timing at CDF,  K. Einsweiller, E. Paganis, and G. Polesello for discussions on triggering, calorimetery
and timing at ATLAS, A. Haas for discussions on the D0 calorimeter, 
and D. B. Kaplan, A. Nelson, and M. Savage for discussions on nuclear capture.  
We thank D. Amidei, L. Dixon, E. Katz,  A. Kraan,   B. Lillie, 
M. Peskin,  J. Qian, T. Rizzo, M. Strassler, L.-T. Wang, and J. Wells for many 
useful conversations.

\end{document}